\documentclass[a4paper,11pt]{article}
\pdfoutput=1 
\usepackage{jcappub} 
\usepackage[T1]{fontenc}
\usepackage{bm,amssymb,amsfonts,slashed,amsthm,amsmath,graphicx}
\graphicspath{ {./Figures/} }

\title{Approximating tunneling rates in multi-dimensional field spaces}

\author[1]{Ali Masoumi,}
\author[1]{Ken D. Olum,}
\author[1]{and Jeremy M. Wachter }

\affiliation[1]{Institute of Cosmology, Department of Physics and Astronomy, 
Tufts University, Medford, MA 02155, USA}

\emailAdd{ali@cosmos.phy.tufts.edu}
\emailAdd{kdo@cosmos.phy.tufts.edu}
\emailAdd{Jeremy.Wachter@tufts.edu}

\def\phitv{{\boldsymbol \phi_{\rm \text{tv}}}}
\def\phifv{{\boldsymbol \phi_{\rm \text{fv}}}}

\newcommand{\bel}[1] {\begin{equation}\label{#1}}
\newcommand{\beal}[1] {\begin{eqnarray}\label{#1}}
\newcommand{\be}{\begin{equation}}
\newcommand{\ee}{\end{equation}}
\newcommand{\bea}{\begin{eqnarray}} 
\newcommand{\eea}{\end{eqnarray}}

\def\({\left(}
\def\){\right)}
\def\[{\left[}
  \def\]{\right]}
\def\BUP{B_{\rm upper}}

\abstract{Quantum mechanics makes the otherwise stable vacua of a theory metastable through the nucleation of bubbles of the new vacuum. This in turn causes a first order phase transition. These cosmological phase transitions may have played an important role in settling our universe into its current vacuum, and they may also happen in future. The most important frameworks where vacuum decay happens contain a large number of fields. Unfortunately, calculating the tunneling rates in these models is very time-consuming. In this paper we present a simple approximation for the tunneling rate by reducing it to a one-field problem which is easy to calculate. We demonstrate the validity of this approximation using our recent code ``Anybubble'' for several classes of potentials.}

\begin{document}
\maketitle
\flushbottom

\section{Introduction} 
\label{sect:Intro}
Quantum tunneling makes the vacua of a field theory metastable. All vacua which are not the global minimum of the effective potential (neglecting gravity) will eventually decay. These decays, which are carried by nucleating bubbles of the new phase, have important cosmological consequences. They may help us to understand how we ended up in our current vacuum, and also may have played an important role in our past. The initial conditions for the inflationary expansion of the universe may have been set by the bubble nucleation that brought us to our current vacuum. Therefore, a good understanding of vacuum decay is of great importance for theoretical physics. 

The problem of vacuum decay in the context of cosmology was first studied by Coleman~\cite{Coleman:1977py}, and the effects of gravity on it by Coleman and De Luccia~\cite{Coleman:1980aw} for the case of a single field with a canonical kinetic term and assuming $O(4)$-symmetric solutions. The only analytic insight comes from the thin-wall approximation. Despite the immense effort and progress over the last few decades, we do not have a clear picture of tunneling in ``realistic'' cases, and past research has not addressed very generic cases, as we explain here.
\begin{itemize}
\item The cases with thin-wall solutions tend to have high actions and hence they are unlikely to have happened in the lifetime of the universe. 
\item If there were tunneling events in our past, gravity probably played important role in it. But, our understanding of the cases with gravity is limited, as we do not know whether or not the calculations based on $O(4)$-symmetric solutions hold in the presence of gravity. There are also important conceptual gaps in our understanding of one of the most important gravitational channels, the Hawking-Moss bounces~\cite{Weinberg:2006pc}.
\item 
  The string landscape is probably the most important application of tunneling formalism. But this landscape has a large number of fields, $N$, and calculation of tunneling rates is a computationally challenging procedure.
  
\item Moreover, the kinetic terms in the string landscape are not canonical, as they are determined by K\"{a}hler metric. It is not very clear whether the formalism developed for theories with canonical kinetic terms is directly applicable to these theories.
\end{itemize}
In this paper, we will only be concerned with the calculation of tunneling rates for cases with a large number of fields. Recently, a very efficient method for calculating the tunneling action in multi-field theories was developed~\cite{Masoumi:2016wot} which uses shooting methods for finding the bounce solutions. This alleviates the problem of calculating tunneling rates for moderately large $N$; however, the run time for $N$ greater than a few can make the computation cost prohibitive when conducting a large survey. Because the decay rates are exponentially sensitive to the tunneling action, if one wants to study the decay rate of a specific vacuum, it is necessary to have the value of tunneling action with a high accuracy. However, in situations where one is interested in understanding the overall distribution of tunneling rates, it suffices to have an estimate of these actions. One example is the surveys done in Refs.~\cite{Greene:2013ida,Masoumi:2016eqo}, where only the median values and distribution of the tunneling actions are calculated. These surveys can help us to understand whether a landscape can generically have long-lived vacua like the one we live in.  

The bubble profile in a multi-dimensional field space traverses the $N$-dimensional field space, following the path with the least action. Restricting the path to a subset of the field space prevents the field from exploring the whole space in the hope of finding a lower action. Since the bounce action always increases as one restricts the bounce to a subset of the field space~\cite{Dasgupta:1996qu}, any bounce restricted in the field space would give an upper bound for the value of the action and a lower bound on tunneling rates.\footnote{A lower bound for tunneling action for polynomial potentials was proposed in \cite{Aravind:2014pva}.}  In this paper we conjecture that if one chooses a particular set of one-dimensional trajectories in the field space and calculates the bounce on them, this upper bound tends to give a very good approximation for the actual value of the action. Our recently developed, efficient ``Anybubble'' code enables us to test this conjecture for several types of landscapes and it appears that our approximation is valid for these important cases. 

We do not claim that our approximations are good always.  In fact,  some potentials have features that make our approximations very  poor.  Nor is there any clear choice of what measure one could choose on  the vast space of possible potentials to claim that the approximations  work a certain fraction of the time.  Instead we have chosen some  particular classes of potentials to test the idea and for those  classes, the approximation works quite well.

 One place where one might make a survey of possible potentials is the landscape of string theory.  This landscape is too large and complex to
 be surveyed, so analyses of landscape models look only in some  particular part of the landscape, and our philosophy here is the same.

 This paper is organized as follows: In Section~\ref{sec:Background} we review the tunneling problem in one- and multi-field cases. In Section~\ref{sec:OneDPot} we propose an approximation for tunneling rates and argue that it should give good results. In Section~\ref{sec:Numerical} we compare our approximation with exact calculations for several important landscapes. In Section~\ref{sec:possible} we explain under what circumstances this approximation may fail. We conclude in Section~\ref{sec:conclusion}. 

\section{Vacuum decay}\label{sec:Background}

Consider a field theory with a potential $U$ having two distinct minima $\phitv$ and $\phifv$. Let us further assume that $U(\phitv)<U(\phifv)$. 

Generally, $\phifv$ will decay into $\phitv$ through a bubble nucleation process if there are no other low minima closer to $\phifv$. In the case that there are other minima which are closer and/or have lower barriers, tunneling occurs to the other vacua and the question of tunneling to $\phitv$ is not very relevant. Assuming the two vacua are close enough to each other, one can calculate the tunneling rate by solving the Euclidean equations of motion~\cite{Coleman:1977py}. For the case of one field it was proved in~\cite{Coleman:1977th} that the solution with the lowest action has an $O(4)$ symmetry. This was generalized recently~\cite{Blum:2016ipp} to the cases with more than one field. Hence the field profile is given by $\phi(x)= \phi(\rho)=\phi(\sqrt{\tau^2+ \mathbf x\cdot \mathbf x})$, given that 
\bel{EOMS}
\frac{d^2\phi_i}{d\rho}+ \frac{3}{\rho}\frac{d\phi_i}{d\rho}=\frac{\partial U}{\partial \phi_i}\,,
\ee
with the boundary conditions 
\bel{BCS}
\phi'_i(0)=0~, \text{ and} \quad \lim_{\rho \rightarrow \infty} \phi(\rho)=\phifv~.
\ee
It is also assumed that $\phi(0)$ is somewhat close to $\phi_{\rm tv}$. The last statement means that $\phi(0)$ should be at a location which can classically roll down to $\phi_{\rm tv}$. If $\phi(\rho)$ is the field profile that satisfies these equations of motion the tunneling action is given by 
\bel{eq:action1}
	S[\phi(\rho)]=2\pi^2 \int_0^\infty d\rho \rho^3\left[\frac12 \partial_\rho\phi_i\partial_\rho\phi_i + U(\boldsymbol\phi(\rho))-U(\phifv)\right]\,.
\ee
For more than one field, it is generic to have more than one solution satisfying the equations \eqref{EOMS} and \eqref{BCS}. These are related to different saddle points of the potential. The tunneling is carried by the instanton which has the least action amongst this set.\footnote{Even for the case of one field, there are situations where the bounce solution is not unique. For more explanation refer to~\cite{Masoumi:2016wot}.} In general performing computation for each calculation of the bounce solution is an expensive computation. Even for a very fast code like ``Anybubble'' it takes several hours to find the tunneling action for a landscape of ten fields. In such a landscape, the number of different instantons responsible for tunneling is also large and hence one computation of the tunneling rate is expensive. If one wants to get an idea about the distribution of the decay rates in a landscape, it usually takes a very large number of bounce calculations, which is too expensive for large $N$ and hence not feasible. However, if we are interested in the overall distribution of the decay rates, we do not need an exact computation and an approximation will suffice. In the next section, we present such an approximation and test its accuracy. 

\section{Reducing to one-dimensional potentials}\label{sec:OneDPot}
The field profile which solves the Euclidean equations of motion traverses a one-dimensional path in the field space. This field profile minimizes the action in directions perpendicular to the field profile. However, it is not a minimum of the (unconstrained) action as it has a negative mode in the direction of expansion-contraction of the bubble. This is why the problem is not amenable to a variational treatment.

\begin{figure}
  \centering
  \includegraphics[width=2.7in]{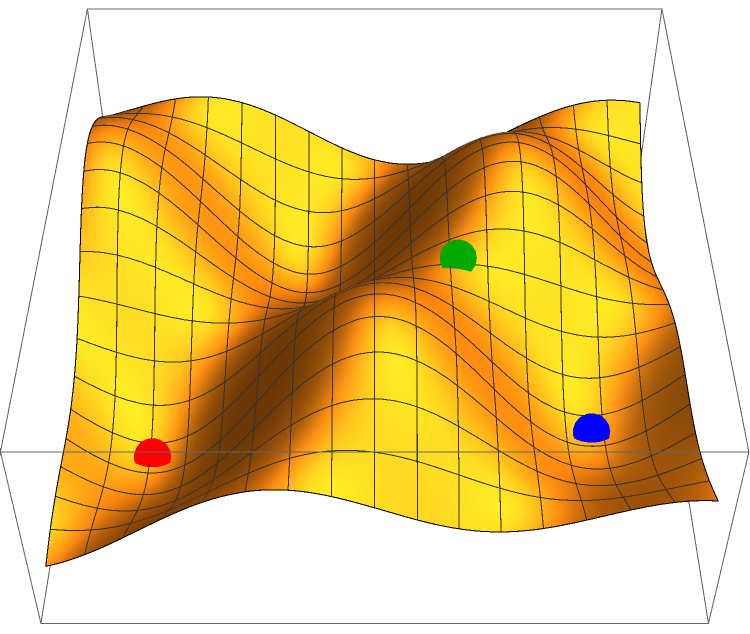}
  \includegraphics[width=2.7in]{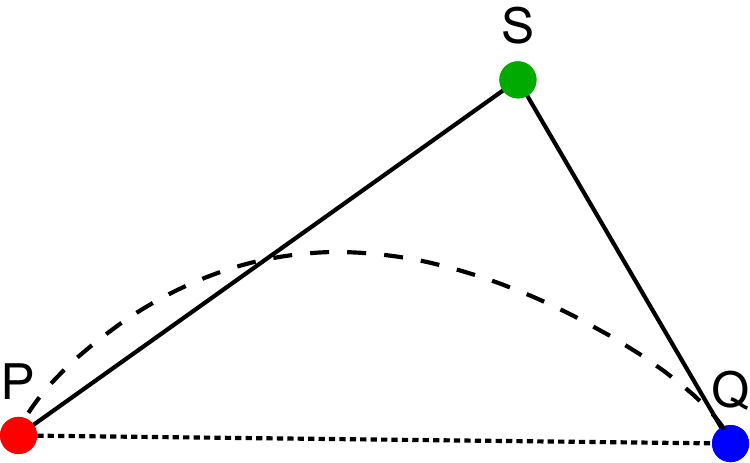} 
   \caption{Left panel, a typical two-dimensional potential. Right panel, solid line is the broken path that can be taken in the field space going from the false vacuum (Q) through the closest saddle point (S) to the true vacuum (P). Dashed line is the trajcetroy that field traverses for the exact bounce and the dotted line is the striaght line that connects the two minima.}
  \label{fig:path1}
\end{figure}

If one constrains the path to an arbitrary subset of the field space, it should have a higher action than the exact action because the profile cannot minimize in these transverse directions. Now, if one chooses an arbitrary path in the field space and treats it as a one-field potential, solving the field equations for the one-dimensional potential along this path should give an upper bound on the action. We conjecture\footnote{This prescription was originally proposed in~\cite{Dasgupta:1996qu} as an upper bound for the action.} that if one chooses a small particular subset of the paths in the field space, shown in Fig.\ref{fig:path1}, this upper bound will be close to the actual value of the action.  We take the one-dimensional potential $U(\lambda)$ along a path parametrized by the path length in the field space.

Intuitively, we expect that tunneling happens through the lowest possible barrier and shortest possible path in the field space. As an example, for the thin-wall cases, the tunneling action is given by 
\bel{thinWall}
B= \frac{27 \pi^2\sigma^4}{2 (U(\lambda=0)-U(\lambda=\lambda_{\rm max}))^3}, \quad \text{where} \quad \sigma= \int_0^{\lambda_{\rm max}} \sqrt{2 U(\lambda)} d\lambda
\ee
where $\lambda$ parameterizes the path length in the field space, starting from the false vacuum. It is clear that a shorter $\lambda$ or a lower $U$ will lead to a smaller action. This suggests two one-dimensional approximations to try: the first is the one with the shortest possible path length, which is a straight line between the two minima, while the second is the path with the lowest possible barrier, in which a ``broken line'' goes through the lowest ridge in the barrier.

The straight line prescription is clear; we explain further the lowest barrier concept in Fig.\ref{fig:path1}. Let us consider a codimension-one family of hyperplanes $P_{\alpha}$ perpendicular to the line connecting the two vacua. Here $\alpha$ parametrizes the planes according to their distances from $\phitv$ with $0 \le \alpha \le |\phitv- \phifv|$.  Because the potential rises moving away from both vacua, we expect to get the lowest ridge to be a maximum in the direction of changing $\alpha$ and a local minimum in the plane $P_\alpha$. These saddle points, whose Hessian has only one negative eigenvalue, are the lowest ridges we are interested in. For example in Fig.\ref{fig:path1} the green dot shows one such point.

For general multi-dimensional field spaces, there will be many bounce solutions with different actions, and the tunneling rate will be dominated by the solution with the lowest action.  To find it requires looking for bounces starting from many different paths.  For example one might try initial conditions going through each saddle point near the true and false vacua.

But we do not attempt to solve this problem.  The purpose of our technique is to find the approximate action of each bounce.  Consider a bounce found by taking a path through saddle point $S$ as the initial condition.  This may be well approximated by the one-dimensional potential on the broken line going through $S$.  However, the tension between the lowest barrier and the shortest path may lead the correct bounce to lie closer to the straight line path then the broken line path.  So one-dimensional potential on the straight line may be a better approximation.  We take the lower of the two actions as our approximation.

\section{Numerical results}\label{sec:Numerical}

We tested the accuracy of the approximation described in previous section for a set of different classes of potentials. All of these potentials were generated with random parameters. For each realization of the potential, we found all the minima and the stationary points whose Hessian had one negative eigenvalue.\footnote{Finding all stationary points of a function numerically is not possible in general.} Now, for each vacuum (local minimum) $\boldsymbol\phi_{\rm min}$, we found the closest minimum $\boldsymbol\phi_{\rm close}$. If this closest minimum had a lower potential, we calculated the tunneling action from $\boldsymbol\phi_{\rm min}$ to $\boldsymbol\phi_{\rm close}$. We calculated this action in three different ways.  First is the action found by solving the $N$-dimensional field equations starting with a path through the closest saddle. Let's call this $B_{\rm exact}$, as it should be the ``true'' action for this class of path. We also calculated the action for the straight-line approximation $B_{\rm straight}$ and the broken-line approximation $B_{\rm broken}$. Since both approximations give upper bounds~\cite{Dasgupta:1996qu} on the action, we kept the smaller of $B_{\rm straight}$ and $B_{\rm broken}$ and called it $\BUP$, after which we calculated the ratio $r=\BUP/B_{\rm exact}$. This ratio should always be more than one (as $B_{\rm exact}$ is the minimum action). We show the distribution of these ratios for different classes of potential in the following subsections.

Sometimes the solution for the straight line path is close to a different bounce from the one with which we are comparing. In that case, we may find $B_{\rm straight} < B_{\rm exact}$, in which case we discard this particular test. This typically happens $2\%$ of the time or less in the tests discussed below; see Appendix~\ref{appendix:straight}. There may also be cases where $B_{\rm straight}$ is greater than but close to $B_{\rm exact}$, because $B_{\rm straight}$ is a poor approximation to a different path with lower action.  We cannot identify these cases , but we think the fraction of them is comparable to the fraction that we discard, and so makes little difference to the overall analysis.

\subsection{Fourier potentials}\label{subsec:Fourier}
\begin{figure}
  \centering
  \includegraphics[scale=0.5]{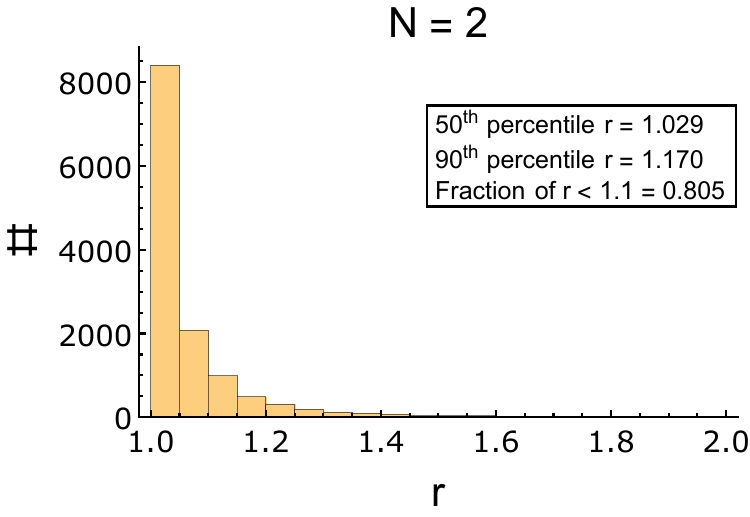}\hspace{5mm}
  \includegraphics[scale=0.5]{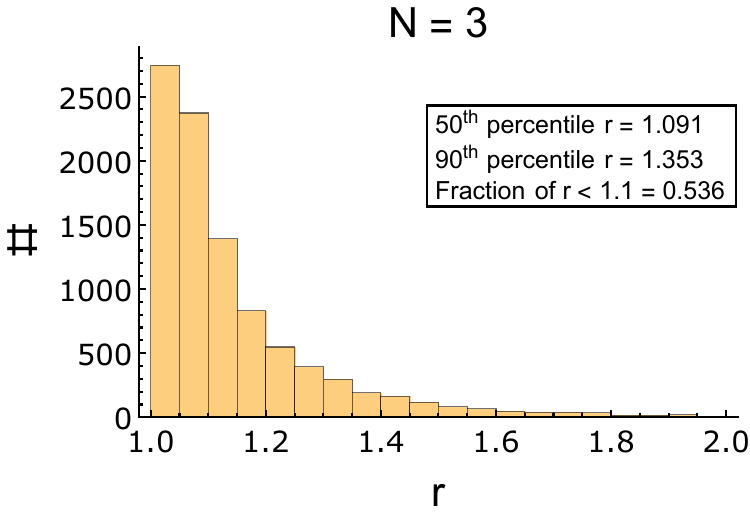}\\ \vspace{5mm}
  \includegraphics[scale=0.5]{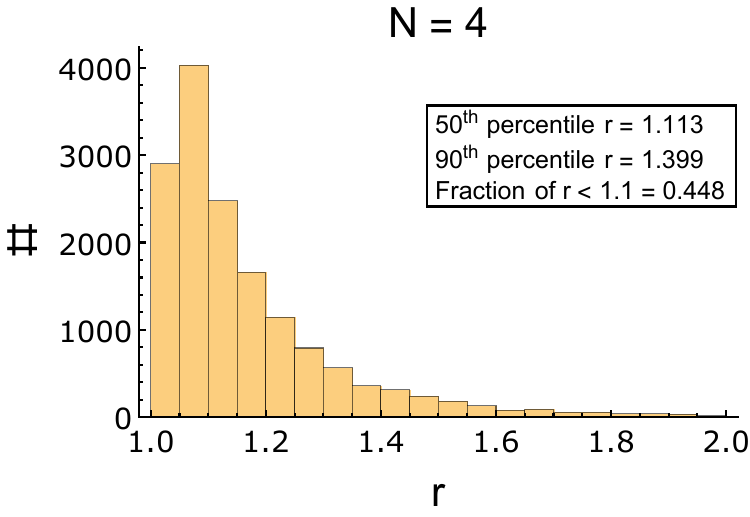}\hspace{5mm}
   \includegraphics[scale=0.5]{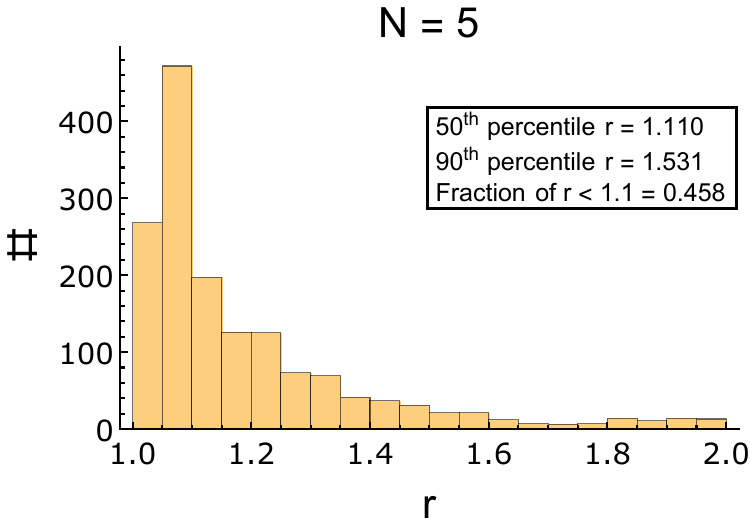}
  \caption{The ratio of the best approximate action to the exact action ($r=\BUP/B_{\rm exact}$) for the Fourier potentials described in \eqref{fourierPot}. Each bar of the histogram is a bin with width $0.05$, and so all results in the two leftmost bars of each plot are for results within ten percent of the exact action.}\label{fig:FourierDist}
\end{figure}

One interesting class of potentials that are of importance are potentials which contain a sum of sine and cosine functions. These potentials appear in many contexts including axionic landscapes and random Gaussian landscapes. We generated a set of potentials in the form\footnote{The problem of tunneling in these landscape was extensively studied in~\cite{Masoumi:2016wot}.}
\bel{fourierPot}
U(\phi)= \sum_{i=1}^{N_c} \alpha_i \cos\(\alpha_i+\sum_{j=1}^{N_{f}} n_{ij} \phi_j\)\,.
\ee
Here, $N_C$ and $N$ are the number of cosine terms in the potential and the number of fields. We chose the coefficient $\alpha_i$ from a Gaussian distribution of zero mean and standard deviation of 0.1. We chose $N_C=15$ and chose the integers $n_{ij}$ from a uniform distribution in the range $[-n_{\rm max}, n_{\rm max} ]$, excluding zero. We chose $n_{\rm max}=6$ except for five fields where we chose $n_{\rm max}=4$ to make the computation tractable. We plotted the histogram of $\BUP/B_{\rm exact}$ in  Fig.\ref{fig:FourierDist}

\subsection{Quartics bounded from below} \label{subsec:Quartic-gen}
Another important class of potentials are quartic polynomials. These are the the highest power polynomials in four dimensions which are renormalizable. So we now consider a class of potentials of the form
\be\label{eqn:poly-pot}
U(\boldsymbol\phi)=\sum_{ijkl}A_{ijkl}\phi^i\phi^j\phi^k\phi^l+\sum_{ijk}B_{ijk}\phi^i\phi^j\phi^k+\sum_{ij}C_{ij}\phi^i\phi^j\,,
\ee
with $N$ fields in total and the only constraint that the resulting quartic be bounded from below. This is achieved by making sure that the quartic terms are ``positive-definite''.   We construct  $A_{ijkl}$ from a family of rank-two matrices $e^{(p)}_{ij}$ as
\be
A_{ijkl}=\sum_{p,q}e^{(p)}_{ij}e^{(q)}_{kl}\,.
\ee
The family of rank-two matrices are obtained by generating a diagonal $N\times N$ matrix with random positive eigenvalues drawn from a uniform distribution, then applying a random $O(N)$ transformation and thereby leaving the eigenvalues unchanged.  The $C_{ijk}$ and $B_{ij}$ matrices have all elements random and drawn from a uniform distribution.  By shifting the origin to a local minimum of the potential, we omitted the need for linear terms in Eq.~\eqref{eqn:poly-pot}. Of course because these potentials are bounded from below there is a global minimum and hence we can always find such a local minimum. We did not include a constant term as the tunneling rates are insensitive to changes in height.

\begin{figure}
  \centering
  \includegraphics[scale=0.5]{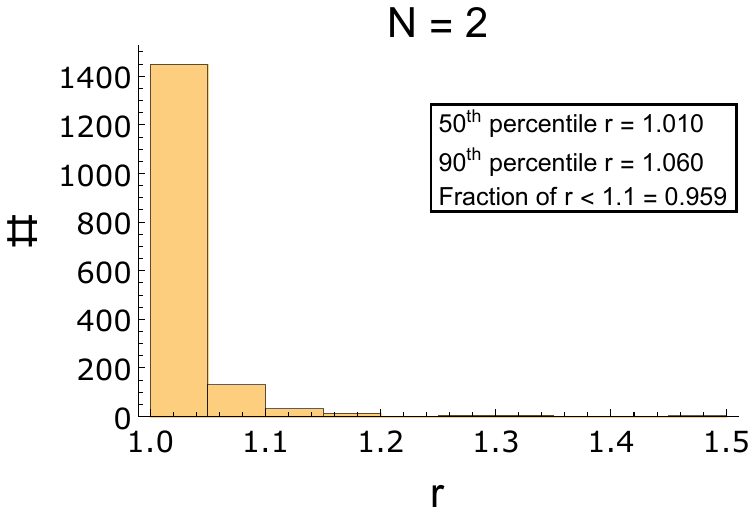}\hspace{5mm}\includegraphics[scale=0.5]{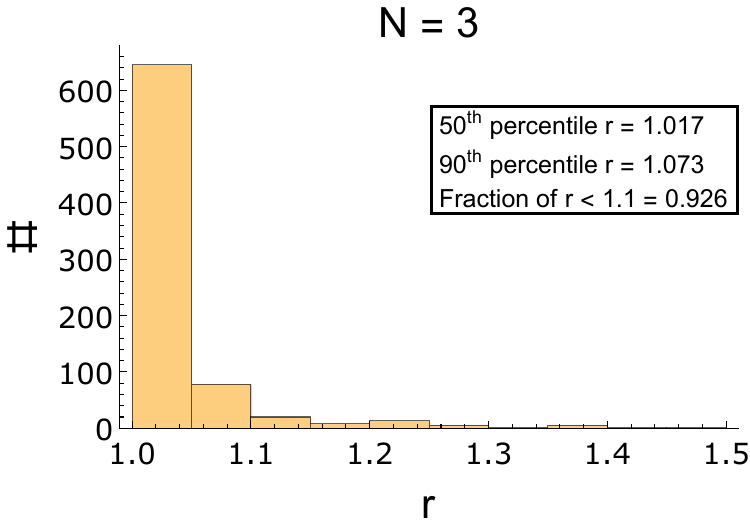}\\ \vspace{5mm}
  \includegraphics[scale=0.5]{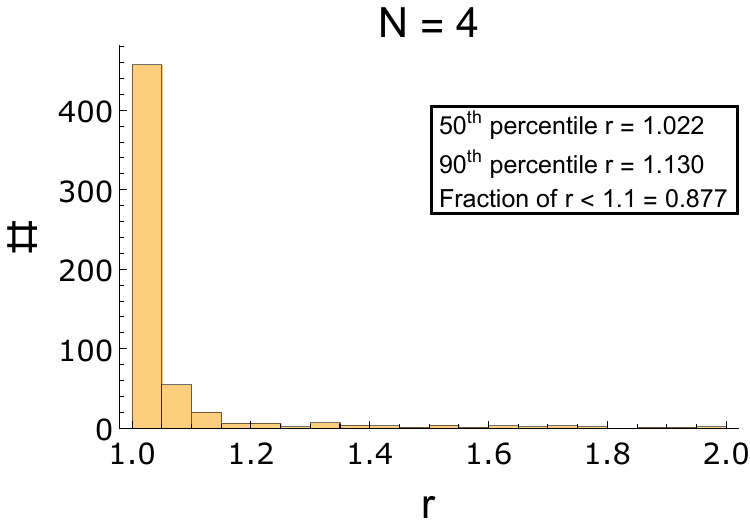}\hspace{5mm}\includegraphics[scale=0.5]{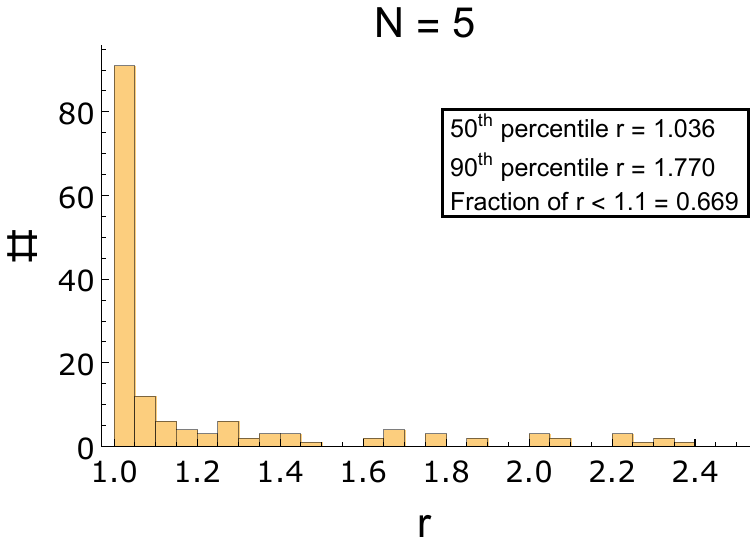}
  \caption{The ratio of the best approximate action to the exact action ($r=\BUP/B_{\rm exact}$) for quartics bounded from below, as described in \eqref{eqn:poly-pot}. Each bar of the histogram is a bin with width $0.05$, and so all results in the two rightmost bars of each plot are for results within ten percent of the exact action.}\label{fig:quartic-gen-hist}
\end{figure}

We report the results of our computations for $N=2,3,4$, and $5$ fields in Fig.~\ref{fig:quartic-gen-hist}. 
These results exhibit some decay in the accuracy of the approximations as the number of fields increases, but the majority of $\BUP$ are always within ten percent of the exact action. The observed decay is consistent with our earlier discussion about perpendicular hyperplanes of codimension one. As the total number of dimensions increases, so too does the number of unoptimized directions in a one-dimensional approximation.

\section{Some possible issues}\label{sec:possible}
In this section we explain some issues about our procedure and also cases where the approximation may fail. We had a success rate varying from 95\% to 80\% for finding the exact bounce. In the cases that the code could not find the exact solution, we could not report the result. Although we believe it is unlikely, if there was correlation between the cases that the approximation is poor and the cases where the code fails, our statistics would be biased.

Another issue is that, as mentioned above, the tunneling solution in multi-field cases (and even in one-field cases) is not necessary unique. Hence, our code may have converged to one solution which was not close to the saddle point where it was intended to converge. This can again bias the statistics and in some cases the exact solution had higher action. We omitted these cases as we know the exact action must have lower action and the fact that its action was higher means that we were looking at a different solution.

Our approximation seems to work well within less than 20 percent error in most cases. However, it is possible that there are cases where there is a bounce solution, but the one-dimensional potentials we construct do not allow for a solution. Although we believe these are very rare, one should be careful in such cases. We discuss these cases in Appendix~\ref{appendix:failure}

\section{Conclusions}\label{sec:conclusion}

Calculation of tunneling action in theories with many fields proves to be a daunting numerical problem. Although there are some codes capable of calculating these actions efficiently, surveys with a large number of fields are not usually tractable. In a large field space with many fields, there will be many bounces (instantons) that carry the tunneling. Calculating even one of these tunneling rates can take a long time, let alone calculating all possible bounces and choosing the smallest of these actions. If one is interested in finding the decay rate of a specific vacuum, there is no viable alternative to going through this process. As the decay rates are exponentially sensitive to the actions, there is no shortcut or approximation which can capture the rate correctly. However, if all one wants is to find the distribution of tunneling rates, having an approximation of tunneling rates will be very useful. We have shown that an upper bound which was proposed in the past can be used as a good approximation for tunneling actions. We checked this conjecture for two generic landscapes, a Fourier landscape and a polynomial landscape, and showed that for most cases this gives the action within $10\%$ of the actual value. Having this approximation, which essentially converts the tunneling problem to a one-dimensional case, can pave the way for calculating the distribution of the tunneling rates for large landscapes. This approximation was tested in the absence of gravity, but it is plausible that it extends to the cases which include gravity.

\section{Acknowledgements}

We are thankful to Alex Vilenkin and Erick Weinberg for useful comments and discussion. This work was supported in part by the National Science Foundation [grant number PHY-1518742] and the John F. Burlingame Graduate Fellowship. The computation was done on the Tufts Linux Research Cluster.

\appendix

\section{One-dimensional potentials with no bounce solution}\label{appendix:failure}

\begin{figure}
  \centering
  \includegraphics[scale=0.5]{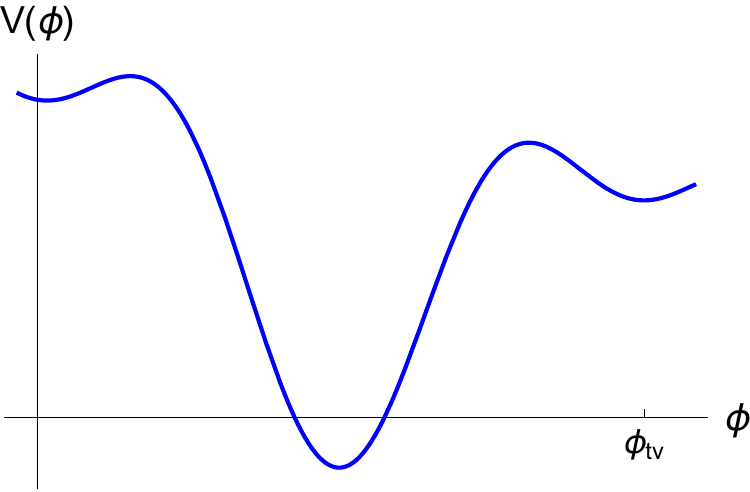}\hspace{5mm}\includegraphics[scale=0.5]{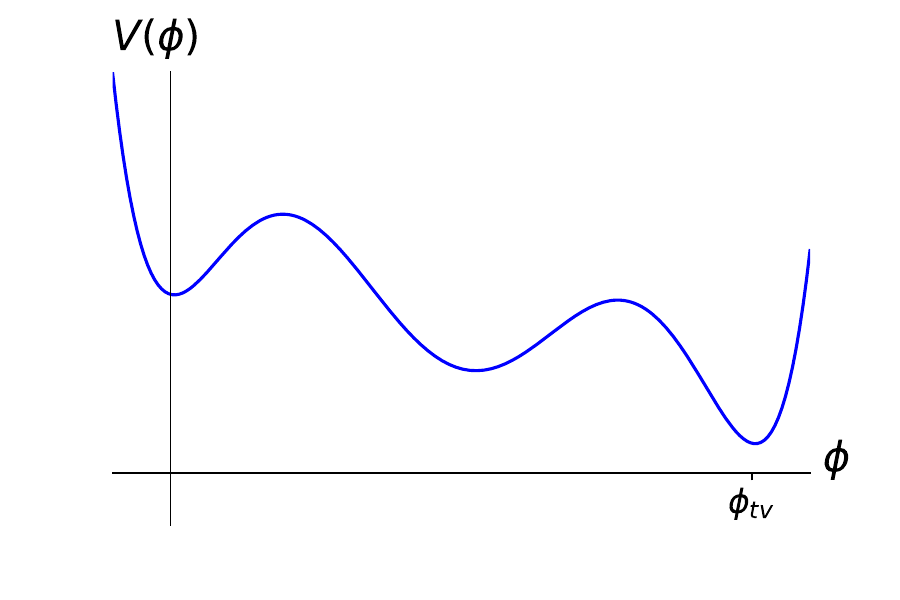}\\ \vspace{5mm}
  \caption{Two examples of one-dimensional potentials where there are no bounces for tunneling from the metastable vacuum at the origin to the true vacuum at $\phi_{\rm tv}$. The left case is easy to understand as the real tunneling happens to the vacuum in between the two vacua. The right hand side case is more interesting as the overshoot-undershoot argument does not work.}
\label{fig:noSol}
\end{figure}

There may be cases where the full $N$-dimensional potential allows for a bounce solution, but there are no one-dimensional solutions along the one-dimensional paths we chose. We believe these cases are rare as we construct the one-dimensional potential along a path to the closest saddle point and closest minimum,  and hence we do not expect any strange feature. But if the one-dimensional potential develops local minima, there may be no one-dimensional solution. We show two such examples of one-dimensional potentials in Fig.\ref{fig:noSol} where there are no tunnelings from the vacuum at the origin to the one at $\phi_{\rm tv}$. The left case is easy to understand as the tunneling happens to the vacuum in between the two vacua. The right side is more interesting. The overshoot-undershoot argument does not work. If the bounce can overshoot the first barrier, it will always overshoot. 

\section{Straight approximations to different exact actions}\label{appendix:straight}
As discussed in Sec.~\ref{sec:Numerical}, there are cases when the straight-line approximation is close to some bounce which is not the bounce for which we find $B_{\rm exact}$. Such cases may have $B_{\rm straight}<B_{\rm exact}$, although this is by no means required. However, since we can definitively identify these cases, they serve as a measure for the frequency with which the straight-line approximation ``belongs'' to a different bounce. The percentages of calculations which reported $B_{\rm straight}<B_{\rm exact}$ are given in Table~\ref{tbl:straight}, grouped by field number and landscape type.
\begin{table}
\centering
\begin{tabular}{|r|r|r|}
\hline
$N$ & Fourier & Polynomial \\
\hline
2 & $1.06\%$ & $0.48\%$ \\
3 & $1.78\%$ & $0.00\%$ \\
4 & $2.12\%$ & $0.00\%$ \\
5 & $11.76\%$ & $0.65\%$ \\
\hline
\end{tabular}
\caption{The percentage of calculations wherein $B_{\rm straight}<B_{\rm exact}$, grouped by field number and landscape type. We na\"ively assume that the percentage of calculations in which the straight-line approximation is close to a bounce other than the one we are studying is comparable.}\label{tbl:straight}
\end{table}

While seven of the scenarios have a low percentage of such cases, the $N=5$ Fourier landscape does not. While we excluded such ``bad'' cases from our analyses in Figs.~\ref{fig:FourierDist} and \ref{fig:quartic-gen-hist}, this does raise the concern that $\approx10\%$ of the reported ratios in the $N=5$ Fourier histogram are in fact approximations to a different bounce.

\bibliography{TunnelingApprox}
\end{document}